\documentclass[12pt]{article}
\usepackage{graphicx}
\begin{document}
 \noindent {\footnotesize\it Astronomy Letters, 2011 Vol. 37, No. 8, pp. 526--535}

 \noindent
 \begin{tabular}{llllllllllllllllllllllllllllllllllllllllllllll}
 & & & & & & & & & & & & & & & & & & & & & & & & & & & & & & & & & & & & & \\\hline\hline
 \end{tabular}

 \vskip 1.5cm
 \centerline {\large\bf Galactic Kinematics from OB3 Stars with Distances}
 \centerline {\large\bf Determined from Interstellar Ca II Lines}
 \bigskip
 \centerline {V.V. Bobylev$^{1,2}$, and A.T. Bajkova$^1$}
 \bigskip

{\small\it
 $^1$~Pulkovo Astronomical Observatory, Russian Academy of Sciences, Pulkovskoe sh. 65, St. Petersburg,
 196140 Russia; E-mail: vbobylev@gao.spb.ru

 $^2$~Sobolev Astronomical Institute, St. Petersburg State University, Universitetskii pr. 28, Petrodvorets, 198504 Russia

 }

 \bigskip

{\bf Abstract}---Based on data for 102 OB3 stars with known proper
motions and radial velocities, we have tested the distances
derived by Megier et al. from interstellar Ca II spectral lines.
The internal reconciliation of the distance scales using the first
derivative of the angular velocity of Galactic rotation
$\Omega'_0$ and the external reconciliation with Humphreys's
distance scale for OB associations refined by Mel'nik and Dambis
show that the initial distances should be reduced by
$\approx$20\%. Given this correction, the heliocentric distances
of these stars lie within the range 0.6--2.6 kpc. A kinematic
analysis of these stars at a fixed Galactocentric distance of the
Sun, $R_0=8$~kpc, has allowed the following parameters to be
determined:
 (1) the solar peculiar velocity components
 $(U_\odot,V_\odot,W_\odot)=(8.9,10.3,6.8)\pm(0.6,1.0,0.4)$ km s$^{-1}$;
 (2) the Galactic rotation parameters
 $\Omega_0=-31.5\pm0.9$~km s$^{-1}$ kpc$^{-1}$,
 $\Omega'_0=+4.49\pm0.12$~km s$^{-1}$ kpc$^{-2}$,
 $\Omega''_0=-1.05\pm0.38$~km s$^{-1}$ kpc$^{-3}$
 (the corresponding Oort constants are
 $A=17.9\pm0.5$~km s$^{-1}$ kpc$^{-1}$,
 $B=-13.6\pm1.0$~km s$^{-1}$ kpc$^{-1}$ and the circular rotation
velocity of the solar neighborhood is
 $|V_0|=252\pm14$~km s$^{-1}$);
 (3) the spiral density wave parameters, namely:
 the perturbation amplitudes for the radial and azimuthal velocity
components, respectively,
 $f_R = -12.5\pm1.1$~km s$^{-1}$ and
 $f_\theta=2.0\pm1.6$~km s$^{-1}$;
 the pitch angle for the two-armed spiral
pattern $i=-5.3^\circ\pm0.3^\circ,$ with the wavelength of the
spiral density wave at the solar distance being
$\lambda=2.3\pm0.2$~kpc; the Sun's phase in the spiral wave
$\chi_\odot=-91^\circ\pm4^\circ$.

 \bigskip

\section*{INTRODUCTION}

Data on various objects are used to determine the Galaxy's
kinematic parameters. These include the radial velocities of
neutral and ionized hydrogen with the distances derived by the
tangential point method (Burton 1971; Clemens 1985; Fich et al.
1989), Cepheids with the distance scale based on the
period--luminosity relation, open star clusters and OB
associations with photometric distances (Mishurov and Zenina 1999;
Rastorguev et al. 1999; Dambis et al. 2001; Zabolotskikh et al.
2002; Bobylev et al. 2008; Mel'nik and Dambis 2009), and maser
sources with trigonometric parallaxes (Reid et al. 2009; McMillan
and Binney 2010; Bobylev and Bajkova 2010).

Young massive luminous stars with large heliocentric distances are
important for the solution of our problem. However, among, for
example, the O-type stars from the Hipparcos (1997) catalog, only
12 have parallaxes differing significantly from zero and the most
distant of them, 10 Lac, is at a distance of only $\approx$540 pc
from the Sun (Maiz-Apell\'aniz et al. 2008). The distances to a
large number of OB stars determined spectroscopically from the
broadening of interstellar CaII, NaI, or KI absorption lines are
of indubitable interest. The method for determining such distances
has long been known (for a review, see Megier et al. 2005).
However, only recently did Megier et al. (2005) and Megier et al.
(2009) tie the ``calcium'' scale, respectively, to the Hipparcos
(1997) trigonometric parallaxes. Megier et al. (2009) estimated
the accuracy of an individual distance to OB stars to be
$\approx15\%.$ For many of them, such highly accurate distance
estimates have been obtained for the first time, given that their
Hipparcos trigonometric parallaxes are not significant. The method
is based on the assumption about a uniform distribution of ionized
atoms in the Galactic plane. As these authors point out, the
derived distances are applicable only at low elevations of stars
above the Galactic plane ($|z|<0.8$~kpc). They also point to
possible local inhomogeneities in Galactic longitude, especially
in the region of the cluster Tr~16.

The best-studied inhomogeneities are those in the immediate solar
neighborhood associated with the Local Bubble. Here, the
variations in the number density of CaII ions reach one order of
magnitude (at a mean number density $n_{CaII}\sim10^{-9}$
cm$^{-3})$ and the spatial sizes of the inhomogeneities are
$\approx$60 pc (Welsh et al. 2010).

Our goal is to test the distance scale derived from interstellar
CaII spectral lines and, subsequently, to use it to investigate
the kinematics of OB stars, in particular, to construct the
Galactic rotation curve and to determine the spiral density wave
parameters. To refine the ``calcium'' scale, we apply the method
for internal reconciliation of the distance scales using the first
derivative of the angular velocity of Galactic rotation
$\Omega'_0$ (Zabolotskikh et al. 2002) and the method for external
reconciliation with Humphreys's (1978) distance scale for OB
associations refined by Mel'nik and Dambis (2009).

\section*{DATA}

We used data on 290 young OB3 stars whose distances were
determined by Megier et al. (2009) from the equivalent widths of
CaII K and CaII H lines by tying them to the trigonometric
parallaxes of a revised version of the Hipparcos catalog (van
Leeuwen 2007). We supplemented the sample with data from the
CRVAD-2 compilation (Kharchenko et al. 2007), with contains the
radial velocities, proper motions, and photometric characteristics
for $\approx$55000 stars.

For spectroscopic binaries, we checked them against the SB9
database (Pourbaix et al. 2004) in order to refine their systemic
radial velocities $V_\gamma$. For a number of stars, we made
significant changes to the CRVAD-2 radial velocities. There are
also the most recent $V_\gamma$ determinations, for instance, for
such stars of our sample as HIP~31978 (Cvetkovi\'c et al. 2010),
HIP~32067 (Mahy et al. 2010), or HIP~78401 (Tango et al. 2009).

As a result, we obtained a working sample of 258 Hipparcos OB3
stars with the distances from Megier et al. (2009), proper motions
(van Leeuwen 2007), and radial velocities. These stars have
various luminosity classes. Note that $\approx20\%$ of the sample
are either known runaway stars (Tetzlaff et al. 2011) or
candidates for runaway stars due to their large ($>40$~km
s$^{-1}$) residual space velocities.

\section*{THE METHOD}

The method used here to determine the kinematic parameters
consists in minimizing a quadratic functional $F$:
$$
\displaylines{\hfill
 \min~F=\sum_{j=1}^N w_r^j
 (V_r^j-\hat{V}_{r}^j)^2+
 \sum_{j=1}^N w_l^j (V_l^j-\hat{V}_{l}^j)^2+\sum_{j=1}^N w_b^j (V_b^j-\hat{V}_{b}^j)^2
\hfill\llap(1)
 }
 $$
where $N$ is the number of stars used; $j$ is the current star
number; $V_r$ is the radial velocity; $V_l = 4.74r\mu_l \cos b$
and $V_b = 4.74r\mu_b$ are the proper motion velocity components
in the $l$ and $b$ directions, respectively, with the coefficient
4.74 being the quotient of the number of kilometers in an
astronomical unit and the number of seconds in a tropical year;
$\hat{V}_{r}^j, \hat{V}_{l}^j, \hat{V}_{b}^j$ are the measured
components of the velocity field (data); $w_r^j, w_l^j, w_b^j$ are
the weight factors, provided that the following constraints
derived from Bottlinger's formulas (Ogorodnikov 1965) with an
expansion of the angular velocity of Galactic rotation $\Omega$
into a series to terms of the second order of smallness with
respect of $r/R_0$ and with allowance made for the influence of
the spiral density wave hold:
$$
\displaylines{\hfill
 V_r=-u_\odot\cos b\cos (l-l_0)-
     v_\odot\cos b\sin (l-l_0)
     -w_\odot\sin b-
 \hfill\llap(2)\cr\hfill
 -R_0(R-R_0)\sin (l-l_0)\cos b \Omega'_0-
 \hfill\cr\hfill
 -0.5R_0 (R-R_0)^2 \sin (l-l_0)\cos b \Omega''_0+  
 \hfill\cr\hfill
 +\tilde{v}_\theta \sin(l-l_0+\theta)\cos b -
 \tilde{v}_R \cos(l-l_0+\theta)\cos b ,
 \hfill\cr\hfill
 V_l= u_\odot\sin (l-l_0) - v_\odot\cos (l-l_0) -
 \hfill
 \cr\hfill
  -(R-R_0)(R_0\cos (l-l_0)-r\cos b) \Omega'_0-
 \hfill\cr\hfill
  -(R-R_0)^2 (R_0\cos (l-l_0)-r\cos b)0.5\Omega''_0 + r \Omega_0 \cos b +
 \hfill\cr\hfill
 + \tilde{v}_\theta \cos(l-l_0+\theta) +
 \tilde{v}_R \sin(l-l_0+\theta),
 \hfill\cr\hfill
 V_b=u_\odot\cos (l-l_0) \sin b +
         v_\odot\sin (l-l_0) \sin b
         -w_\odot\cos b +
  \hfill\cr\hfill
 +R_0(R-R_0)\sin (l-l_0)\sin b\Omega'_0+
  \hfill\cr\hfill
 +0.5R_0(R-R_0)^2\sin (l-l_0)\sin b\Omega''_0- 
 \hfill\cr\hfill
 -\tilde{v}_\theta \sin(l-l_0+\theta)\sin b +
 \tilde{v}_R \cos(l-l_0+\theta)\sin b , \hfill
 }
$$
where $r$ is the star's heliocentric distance; the star's proper
motion components $\mu_l \cos b$ and $\mu_b$ are in mas yr$^{-1}$
and the radial velocity $V_r$ is, in km s$^{-1}$;
$u_\odot,v_\odot,w_\odot$ are the stellar group velocity
components relative to the Sun taken with the opposite sign (the
velocity $u$ is directed toward the Galactic center; $v$ is in the
direction of Galactic rotation; and $w$ is directed to the north
Galactic pole); $R_0$ is the galactocentric distance of the Sun;
$R$ is the galactocentric distance of the star; $l_0$ is the
direction to the kinematic center (to the Galactic center) --- we
included it as an unknown to reveal possible peculiarities of the
``calcium'' stellar distance scale. $\Omega_0$ is the angular
velocity of rotation at the distance $R_0$; the parameters
$\Omega'_0$ and $\Omega''_0$ are, respectively, the first and
second derivatives of the angular velocity; the distance $R$ is
calculated from the expression
$$
\displaylines{\hfill
 R^2=(r\cos b)^2-2R_0 r\cos b\cos (l-l_0)+R^2_0.\hfill\llap(3)
 }
$$
To take into account the influence of the spiral density wave, we
used the simplest kinematic model based on the linear theory of
density waves by Lin et al. (1969), in which the potential
perturbation is in the form of a traveling wave. Then,
$$
\displaylines{\hfill
 \tilde{v}_R = f_R \cos \chi,    \hfill\llap(4)\cr\hfill
 \tilde{v}_\theta = f_\theta \sin \chi,\hfill\cr\hfill
 \chi= m [\cot (i) \ln (R/R_0) - \theta] + \chi_\odot,\hfill
 }
$$
where $f_R$ and $f_\theta$ are the perturbation amplitudes for the
radial (directed toward the Galactic center in the arm) and
azimuthal (directed along the Galactic rotation) velocities; $i$
is the spiral pitch angle ($i<0$ for winding spirals); $m$ is the
number of arms, with $m=2$ taken here; $\theta$ is the position
angle of the star (measured in the direction of Galactic
rotation); $\chi_\odot$ is the phase angle of the Sun, measured
here from the center of the Carina--Sagittarius spiral arm
$(R\approx7$~kpc), as was done by Rohlfs (1977). The parameter
$\lambda$ is the distance (in the Galactocentric radial direction)
between adjacent segments of spiral arms in the solar neighborhood
(the wavelength of the spiral wave)---it is calculated from the
relation
$$
\displaylines{\hfill
 \tan (i) = {{\lambda m}\over {2\pi R_0}}.\hfill\llap(5)
 }
$$
The described method of allowance for the influence of the spiral
density wave was applied by Mishurov and Zenina (1999) and
Fern\'andez et al. (2001), where its detailed description can be
found, and by Zabolotskikh et al. (2002).

The weight factors in functional (1) are assigned according to the
following expressions (for simplification, we omit the index $j$):
$$
\displaylines{\hfill
 w_r=S_0/\sqrt {S_0^2+\sigma^2_{V_r}},
     \hfill\cr\hfill
 w_l=\beta^2 S_0/\sqrt {S_0^2+\sigma^2_{V_l}},
     \hfill\cr\hfill
 w_b=\gamma^2 S_0/\sqrt {S_0^2+\sigma^2_{V_b}},
     \hfill
 }
$$
where $S_0$ denotes the dispersion averaged over all observations,
which has the meaning of a ``cosmic'' dispersion taken to be 8 km
s$^{-1}$; $\beta=\sigma_{V_r}/\sigma_{V_l}$ and
$\gamma=\sigma_{V_r}/\sigma_{V_b}$ are the scale factors that we
determined using data on open star clusters (Bobylev et al. 2007),
$\beta=1$ and $\gamma=2$. The errors of the velocities $V_l$ and
$V_b$ are calculated from the formula
$$
\displaylines{\hfill
 \sigma_{(V_l,V_b)} = 4.74 r
 \sqrt{\mu^2_{l,b}\Biggl({\sigma_r\over r}\Biggr)^2+\sigma^2_{\mu_{l,b}}}.\hfill
 }
$$
The optimization problem (1)--(4) is solved for eleven unknown
parameters
 $u_\odot,v_\odot,w_\odot,\Omega_0,\Omega'_0,\Omega''_0,
 l_0,f_R,f_\theta,i,$ and $\chi_\odot$ by the coordinate-wise descent method (the
sought-for parameters are taken as the coordinates).

We estimated the errors of the sought-for parameters through Monte
Carlo simulations. The errors were estimated by performing 100
cycles of computations. For this number of cycles, the mean values
of the solutions virtually coincide with the solutions obtained
from the input data without any addition of measurement errors.
Measurements errors were added to such input data as the radial
velocities, proper motions, and distances.

Here, we take a fixed value of $R_0.$ Reid (1993) found a weighted
mean from the measurements published over a 20-year period,
 $R_0=8.0\pm0.5$~kpc. Taking into account the main types of errors
and correlations associated with the classes of measurements,
Nikiforov (2003) derived the ``best value'',
 $R_0=7.9\pm0.2$~kpc.
A similar result was obtained by Avedisova (2005),
$R_0=7.8\pm0.3$~kpc. Note several most recent $R_0$
determinations. These include the direct distance measurements
based on the orbits of stars moving around a massive black hole at
the Galactic center, which give $R_0=8.4\pm0.4$~kpc (Ghez et al.
2008) or
 $R_0=8.33\pm0.35$~kpc (Gillessen et al. 2009). A summary of the latest
determinations can be found in Foster and Cooper (2010), where the
weighted mean is $R_0=8.0\pm0.4$~kpc. Given all uncertainties, we
consider $R_0=8.0\pm0.4$~kpc to be the most probable value.

\begin{table}[t]                                                
\caption[]{\small\baselineskip=1.0ex{
 A summary of $\Omega^{'}_0$ and $\lambda/R_0$ determinations}

}
\begin{center}
      \label{t:888}
 \small

\begin{tabular}{|l|l|l|c|l|}\hline
 Reference                  & Method                & Data            & $\Omega^{'}_0 (R_0=8$~kpc), & $~~~~\lambda/R_0$  \\
                            &                       &                 & km s$^{-1}$ kpc$^{-2}$      &               \\\hline
 Clemens (1985)             & $\Delta V_\theta$     & CO+HI           &                    & $0.22$        \\
 Mel'nik et al. (2001)      & $V(\mu)+V_r$          & OB associations &                    & $0.28\pm0.03$ \\
 Zabolotskikh et al. (2002) & $V(\mu)+V_r$          & Cepheids+OSCs   &  $-4.3 \pm0.2$ (*) & $0.33\pm0.04$              \\
 Zabolotskikh et al. (2002) & $V(\mu)+V_r$          & Supergiants     &  $-4.4 \pm0.2$ (*) & $0.36\pm0.05$              \\
 Bobylev et al. (2008)      & $\Delta V_\theta$     & HI+HII+OSCs     &                    & $0.33\pm0.04$ \\
 Bobylev et al. (2008)      & $V_R$                 & OSCs            &                    & $0.23\pm0.07$ \\
 Mel'nik and Dambis (2009)  & $V(\mu)+V_r$          & OB associations &  $-4.4 \pm0.2$     &               \\
 Bobylev and Bajkova (2010) & $V_R$                 & Masers          &  $-4.5 \pm0.2$     & $0.25\pm0.03$ \\\hline

\end{tabular}
\end{center}
{\small
 * : The mean of the values obtained at $R_0=7.5$~kpc and $R_0=8.5$~kpc;
 we calculated $\lambda/R_0$ us from the pitch angle $i$ (for the two-armed model) based on Eq. (5). }
\end{table}

\begin{table}[t]                                                
\caption[]{\small\baselineskip=1.0ex{Kinematic parameters found
using the refined distance scale~($p_{scale}=0.8$)}

}
\begin{center}
      \label{t:999}
 \small
\begin{tabular}{|l|c|c|c|c|c|}\hline

 Parameters                   & No~1            & No~2          & No~3          & No~4          & No~5           \\\hline

 $u_\odot,$    km s$^{-1}$           & $  8.9\pm0.6 $ & $ 8.9\pm0.7$ & $ 9.3\pm0.7$ & $ 9.0\pm0.5$ & $ 9.2\pm0.6$ \\
 $v_\odot,$    km s$^{-1}$           & $ 10.3\pm1.0 $ & $10.3\pm0.9$ & $10.1\pm1.0$ & $10.4\pm1.1$ & $10.2\pm1.0$ \\
 $w_\odot,$    km s$^{-1}$           & $  6.8\pm0.4 $ &          $-$ &          $-$ & $ 6.5\pm0.2$ & $ 6.8\pm0.4$ \\
 $\Omega_0,$   km s$^{-1}$ kpc$^{-1}$   & $-31.5\pm0.9 $ & $-31.5\pm0.8$  & $-31.8\pm1.0$  & $-29.8\pm1.3 $ & $-31.2\pm0.9$ \\
 $\Omega^{'}_0,$ km s$^{-1}$ kpc$^{-2}$ & $ 4.49\pm0.12$ & $ 4.49\pm0.11$ & $ 4.53\pm0.13$ & $ 4.14\pm0.16$ & $ 4.46\pm0.12$ \\
$\Omega^{''}_0,$ km s$^{-1}$ kpc$^{-3}$ & $-1.05\pm0.38$ & $-1.05\pm0.36$ & $-1.15\pm0.42$ & $-0.93\pm0.34$ & $-1.12\pm0.40$ \\
 $f_R,$        km s$^{-1}$    & $-12.5\pm1.1$  & $-12.5\pm0.9$ & $-13.4\pm0.9$ & $ -9.4\pm4.4$ & $-12.7\pm1.1$ \\
 $f_\theta,$   km s$^{-1}$    & $  2.0\pm1.6$  & $  2.1\pm1.5$ & $  0.7\pm1.7$ & $  2.5\pm1.7$ & $  1.8\pm1.6$ \\
 $l_0,$        deg.           & $ -1.5\pm1.3$  & $ -1.5\pm1.2$ & $ -1.1\pm1.3$ & $ -2.3\pm1.2$ &          $-$  \\
 $i,$          deg.           & $ -5.3\pm0.3$  & $ -5.3\pm0.3$ & $ -5.4\pm0.3$ & $ -5.3\pm44$  & $ -5.4\pm0.3$ \\
 $\chi_\odot,$ deg.           & $    -91\pm4$  & $    -91\pm4$ & $    -89\pm3$ & $ -87\pm47$   & $-91\pm4$   \\
 $\lambda,$    kpc            & $  2.3\pm0.2$  & $  2.3\pm0.2$ & $  2.4\pm0.2$ & $   2.3$      & $ 2.4\pm0.2$  \\
 $\sigma_0,$   km s$^{-1}$    &           9.5  &           9.5 &          10.6 &           8.3 &   9.6         \\
 $N_\star$                    &           102  &           102 &           102 &           219 &   102      \\\hline

\end{tabular}
\end{center}
\end{table}

\section*{RESULTS}

When solving the system of equations~(2), 38 of 258 stars are
rejected according to the $3\sigma$ criterion already at the first
step. Seventeen of these stars are identified with the catalog of
candidate runaway stars (Tetzlaff et al. 2011); most of the
remaining ones are either new candidates for runaway stars or have
measurements of the radial velocities or proper motions that are
too unreliable.

Since the Gould Belt has a significant influence in the region
$r<0.8$~kpc, we do not consider any stars from this region when
studying the parameters of the Galactic rotation curve.

There is only one star in the region $r>3.2$~kpc, HIP 85020
($r=4.5$~kpc). It has a low residual velocity, but we do not
consider this star, because it is too far from the common grouping
of stars. As a result, our working sample consists of 102 stars.

\subsection*{The Initial ``Calcium'' Distance Scale}

We obtained the following solution from 102 stars in the range of
distances 0.8--3.2 kpc:
 $(u_\odot,v_\odot,w_\odot,)=(9.3,10.2,8.4)\pm(0.7,1.3,0.4)$ km s$^{-1}$ and
$$
\displaylines{\hfill
 \Omega_0      =-28.6 \pm0.8~\hbox{km s$^{-1}$}, \hfill\llap(6)\cr\hfill
 \Omega^{'}_0  =+3.91\pm0.10~\hbox{km s$^{-2}$}, \hfill\cr\hfill
 \Omega^{''}_0 =-0.81\pm0.35~\hbox{km s$^{-3}$}, \hfill\cr\hfill
         f_R=-11.9\pm1.3~\hbox{km s$^{-1}$},     \quad
      f_\theta=    2\pm2~\hbox{km s$^{-1}$},     \hfill\cr\hfill
        l_0= -1^\circ\pm2^\circ,     \quad
          i=-6.8^\circ\pm0.5^\circ,  \quad
  \chi_\odot=-94^\circ\pm5^\circ,    \hfill
}
$$
the error per unit weight is $\sigma_0=11.2$~km s$^{-1}$. The
almost zero value of $l_0$ shows that there are no significant
global longitudinal peculiarities.

Based on the pitch angle $i$ found in solution (6) and Eq. (5), we
find $\lambda/R_0=0.37\pm0.03$ and then $\lambda=3.0\pm0.2$~kpc.
This value of $\lambda$ is too large compared to the
determinations of this parameter by other authors using various
independent data. This suggests that the ``calcium'' distance
scale is stretched. To determine how much it is stretched, we use
several kinematic methods for comparing the distance scales.

\subsection*{Refining the ``Calcium'' Distance Scale}

We know a method (Dambis et al. 2001; Zabolotskikh et al. 2002)
where the distance scale coefficient $(p_{scale})$ is included as
an additional unknown in the initial kinematic equations (e.g.,
(2)). In this case, it can be determined by simultaneously solving
the system of equations or can be found by minimizing the
residuals according to the $\chi^2$ test.

Unfortunately, this method did not give a reliable result in our
case. The minimum of $\chi^2$ is reached at
$p_{scale}\approx0.3-0.4.$ The new distances are calculated as
$r_{NEW} = p_{scale}\times r.$ However, for such a radical
reduction of the initial ``calcium'' scale, the remaining model
parameters lose any physical meaning. We assume that this is
determined by the ``calcium'' scale calibration method --- the
nonlinear (hyperbolic) relation between the equivalent widths of
CaII spectral lines and trigonometric parallaxes (Megier et al.
2009). Thus, different values of the linear coefficient
$p_{scale}$ should be applied for different ranges of distances.
Other distance scale reconciliation methods are also known.

 \begin{figure}[t]
      {
      \begin{center}
       \includegraphics[width=140mm]{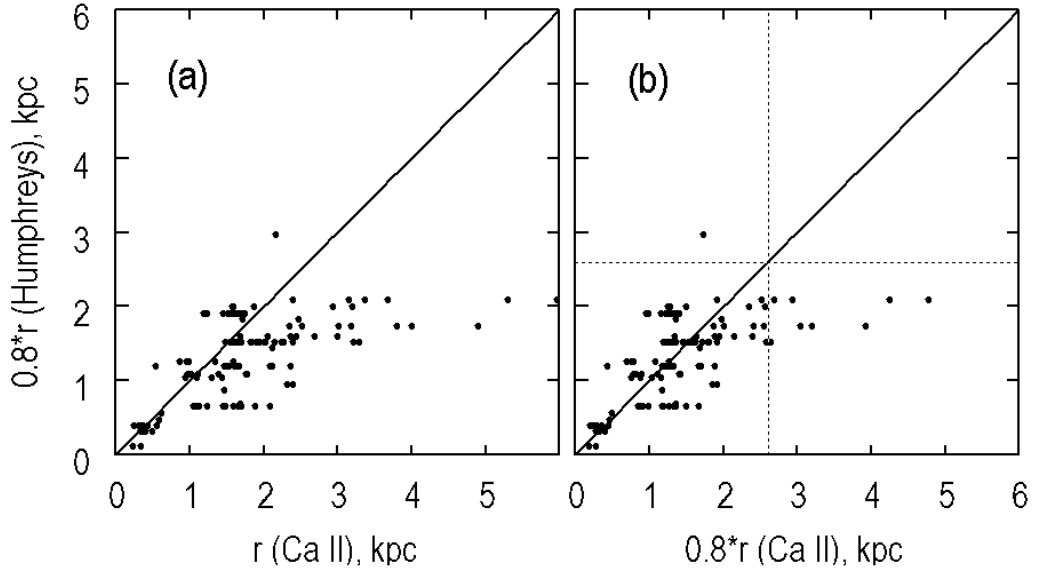}
 \caption{
 (a) Stellar distances $0.8\times r_{Humphreys}$ versus distances $r_{CaII};$
 (b) stellar distances $0.8\times r_{Humphreys}$ versus distances
 $0.8\times r_{CaII};$ the dotted lines mark the outer boundaries of the
sample of stars used to determine the kinematic parameters.}
      \end{center} }
      \end{figure}

\subsubsection*{Reconciliation of the derivative $\Omega^{'}_0.$}

The method for internal reconciliation of the distance scales
using the first derivative of the angular velocity of Galactic
rotation $\Omega^{'}_0$ (Zabolotskikh et al. 2002) consists in
comparing the values obtained by separately solving the system of
equations~(2). For this purpose, we solved a simplified system of
equations with the unknowns $u_\odot,v_\odot,w_\odot$ and
 $\Omega_0,\Omega'_0,\Omega''_0;$ when using the radial velocities,
we fixed the parameters $w_\odot$ and $\Omega_0$ found from the
proper motions. We found
 $\Omega'_0~(Vr)=3.87\pm0.35$ km s$^{-1}$ kpc$^{-2}$ only from the radial velocities and
 $\Omega^{'}_0~(V_\mu)=3.29\pm0.26$~km s$^{-1}$ kpc$^{-2}$ only from the proper motions.
Then, $p_{scale}=3.29/3.87=0.85.$

At present, there are quite satisfactory estimates of
$\Omega^{'}_0$ and $\lambda$ for external scale calibration. A
number of such results are presented in Table 1; they were
obtained in different distance scales: the tangential point method
(Clemens 1985), the Cepheid scale, the photometric scale (open
star clusters --- OSCs) reconciled with the Hipparcos
trigonometric parallaxes, and the trigonometric parallaxes
(masers). By comparing $\Omega^{'}_0=3.9$~km s$^{-1}$ kpc$^{-2}$
found in solution (6) with the mean from Table 2, we obtain
$p_{scale}=3.9/4.4=0.89.$

\subsubsection*{Reconciliation of the wavelength $\lambda$.}

By comparing the mean from Table 1, $\lambda/R_0 = 0.29,$ with the
result of solution (6), $\lambda/R_0=0.37,$ we find
$p_{scale}=0.29/0.37= 0.78.$ Having separately considered the
Galactocentric radial velocities of our OB3 stars, we found
$\lambda/R_0 = 0.36\pm0.03$ $(\lambda= 2.9\pm0.2$ kpc). For this
purpose, we applied the method of Fourier analysis described in
detail previously (Bobylev and Bajkova 2010). In this case,
$p_{scale} = 0.29/0.36 = 0.80.$ Thus, this method gives
$p_{scale}$ close to 0.8.

 \begin{figure}[t]
      {
     \begin{center}
       \includegraphics[width=80mm]{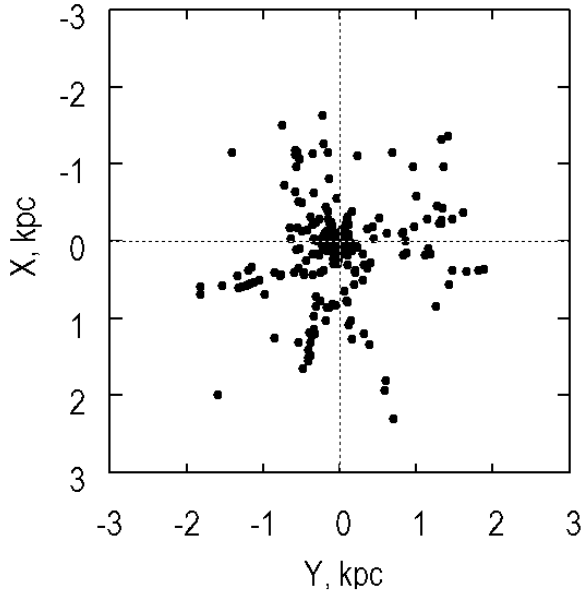}
 \caption{Positions of 219 OB3 stars in projection onto the
Galactic plane.}
   \end{center} }
      \end{figure}

\subsubsection*{Reconciliation with Humphreys's refined distance
scale.}

The membership in OB associations with the distances to them
determined by Humphreys (1978) is specified in the catalog by
Megier et al. (2009) for a considerable number of stars. A number
of researchers of Humphreys's scale (Dambis et al. 2001; Mel'nik
and Dambis, 2009) conclude that it should be reduced by
$\approx20\%,$ i.e., the refined distances should be calculated as
$r_{NEW} = 0.8 \times r_{Humphreys}.$ Figure 1 shows the
``$0.8\times r_{Humphreys}-r_{CaII}$'' and ``$0.8\times
r_{Humphreys}-0.8\times r_{CaII}$'' relations. In fact, the
distance scale by Mel'nik and Dambis (2009) is along the vertical
axis. It is clearly seen from the figure that the ``calcium''
scale should be reduced with a coefficient close to 0.8.

As can be seen from Fig. 1b, the stars with distances greater than
3 kpc already deviate significantly from the general relation even
in the corrected ``calcium'' scale. We did not specially
determined the outer boundary indicated in the figure, because it
was revealed automatically --- the stars outside this boundary
were either rejected according to the $3\sigma$ criterion when
solving the system of equations (2) or had no velocity
measurements.

      \begin{figure}[t]
      {
    \begin{center}
        \includegraphics[width=120mm]{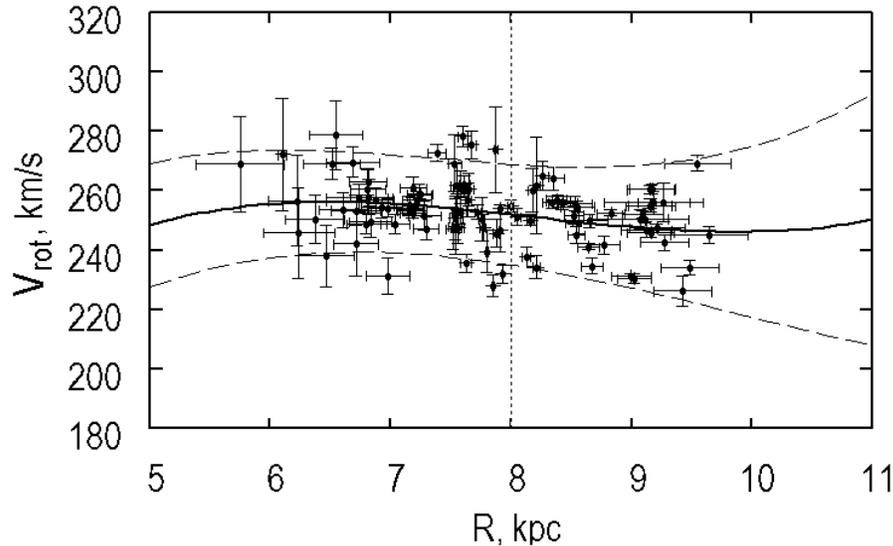}
\caption{Galactic rotation curve (solid line). The vertical line
marks the position of the solar circle. The dashed lines indicate
the $1\sigma$ confidence intervals.}
    \end{center}}
      \end{figure}

      \begin{figure}[t]
      {
      \begin{center}
         \includegraphics[width=110mm]{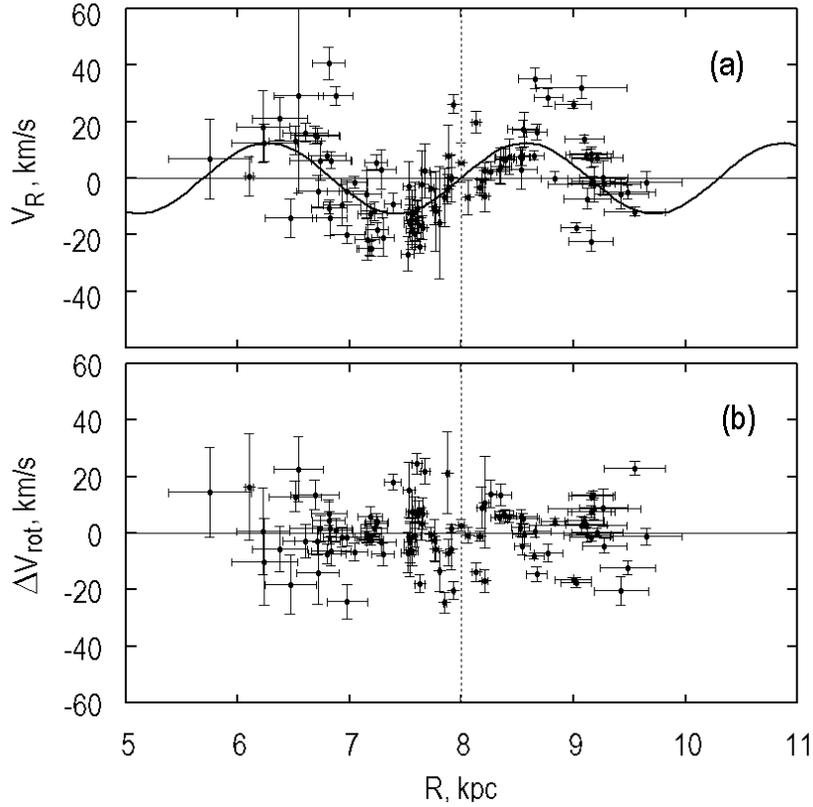}
\caption{Galactocentric radial velocities for 102 relatively
distant OB3 stars (a),
   the sine wave with a period of 2.3 kpc and an
amplitude of 12.5 km s$^{-1}$ (solid line), and the residual
azimuthal velocities (b). The vertical line marks the position of
the solar circle.}
    \end{center}}
      \end{figure}

\subsection*{Galactic Rotation Curve}

Using several methods of analysis, we showed that the scale
coefficient $p_{scale}$ lies within the range 0.7--0.9, with a
mean close to 0.8. We use this mean to form the refined distances
$r_{NEW} = 0.8 \times r_{CaII}$ with which all of the subsequent
computations are performed.

The positions of 219 stars in the Galactic $XY$ plane computed in
the refined distance scale are shown in Fig. 2.

Based on the sample of 102 relatively distant stars located in the
range of distances $r_{NEW}$ from 0.6 to 2.6 kpc, we found several
solutions of the system of equations (2) given in Table 2.
Solution 1 was obtained in the same way as solution (6) using all
tree components $V_r,V_l,V_b.$ To obtain solution 2, we used the
same components $V_r,V_l,V_b$ but fixed the velocity $w_\odot = 7$
km s$^{-1}$. We found solution 3 only from two components, $V_r$
and $V_l,$ with the fixed velocity $w_\odot = 7$ km s$^{-1}$.
Solution 4 was obtained in the same way as solution 1 using all
three components $V_r,V_l,V_b$ from all 219 stars.

Solution 4 shows that the influence of nearby stars degrades
considerably the accuracy of determining the spiral-structure
parameters. Solutions 1 and 2 are virtually identical. Comparison
of solution 3 with solutions 1 and 2 shows that using the three
components $V_r,V_l,V_b$ is more preferable, i.e., the stars are
not too far from the Sun for valuable information to be lost in
the components $V_b.$

Solution 5 was obtained in the same way as solution 1 using all
three components $V_r,V_l,V_b$, but the direction to the Galactic
center was fixed at $l_0 = 0^\circ$. All of the determined
parameters in solution 5 agree well with the results of solutions
1 and 2. Therefore, below we will use the result of the more
general case (solution 1).

Figure 3 displays the Galactic rotation curve constructed
according to solution 1 from Table 2. The dashed lines indicate
the boundaries of the confidence intervals corresponding to a
$1\sigma$ error level calculated by taking into account the
contributions from the error in the angular velocity and the
uncertainty in $R_0.$

Figure 4 presents the Galactocentric radial velocities $V_R$ (the
direction away from the Galactic center is considered the positive
one) for 102 relatively distant OB3 stars and the residual
azimuthal velocities $\Delta V_{rot}$. The azimuthal velocities
are residual, because the Galactic rotation curve found was
excluded from them; both velocities $V_R$ and $\Delta V_{rot}$
were freed from the group velocity. The sine wave associated with
the influence of the spiral density wave was fitted into the
radial velocities (based on solution 1 from Table 2).

The Oort constants calculated using solution 1 from Table 2 are
 $A=17.9\pm0.5$ km s$^{-1}$ kpc$^{-1}$ and
 $B=-13.6\pm1.0$ km s$^{-1}$ kpc$^{-1}$. The
circular rotation velocity of the solar neighborhood is
 $|V_0|=252\pm14$~km s$^{-1}$ (calculated by taking into account the error in $R_0$ of
0.4 kpc).

Table 3 lists the stars with residual space velocities
 $|V_{pec}|>40$~km s$^{-1}$. We calculated the velocities $V_{pec}$ by excluding the
rotation curve found and the solar peculiar velocity as well as by
excluding the wave found in the radial velocities. These stars are
either already known runaway stars or suitable candidates for
runaway stars.

\begin{table}[p]                             
 \caption[]{\small\baselineskip=1.0ex
 {Candidates for runaway stars}

}
\begin{center}
      \label{t:runaway}
\begin{tabular}{|r|c|c|}\hline
   HIP   & $|V_{pec}|,$~km s$^{-1}$ & Remark \\\hline

  18350  &  63$\pm$11  & R \\
  18614  &  54$\pm$15  & R \\
  22783  &  58$\pm$12  &   \\
  24575  & 115$\pm$12  & R \\
  27204  &  97$\pm$3   & R \\
  29147  & 101$\pm$28  &   \\
  31348  &  61$\pm$19  &   \\
  32067  &  40$\pm$6   & R \\
  43158  &  58$\pm$14  & R \\
  54475  &  63$\pm$29  & R \\
  63170  &  59$\pm$15  & R \\
  64896  &  54$\pm$21  &   \\
  65129  &  42$\pm$7   &   \\
  78310  &  46$\pm$9   &   \\
  81100  &  44$\pm$6   & R \\
  81305  &  46$\pm$10  & R \\
  81696  &  68$\pm$125 & R \\
  81702  &  48$\pm$26  &   \\
  82171  &  71$\pm$10  & R \\
  82685  &  41$\pm$20  &   \\
  82775  & 110$\pm$8   & R \\
  83499  &  65$\pm$10  &   \\
  84687  &  48$\pm$37  & R \\
  85331  & 113$\pm$22  & R \\
  89218  &  56$\pm$18  &   \\
  89750  &  78$\pm$40  &   \\
  98863  &  48$\pm$21  &   \\
 100287  &  48$\pm$22  &   \\
 101186  &  42$\pm$6   & R \\
 109556  &  86$\pm$16  & R \\
 114695  &  42$\pm$23  &   \\
\hline
 \end{tabular}\end{center}
 {\small
 Note. R denotes a runaway star according to Tetzlaff
et al. (2011).  }
    \end{table}

\section*{DISCUSSION}

(1) The phase $\chi_\odot$ we found is almost equal to $-\pi/2.$
Its value does not depend on the distance scale. The specific
values of the solar peculiar velocity relative to the local
standard of rest (LSR) depend on the phase $\chi_\odot$ (see Eqs.
(19),(20) in Bobylev and Bajkova 2010). Since young stars
experience perturbations from the spiral density wave, the
components of their mean motion $u_\odot$ and $v_\odot$ can differ
significantly from the velocities found from older stars (Dehnen
and Binney 1998). Denote the solar peculiar velocity components
unperturbed by the spiral wave by
$(U_\odot,V_\odot,W_\odot)_{LSR}.$ As is clearly seen from
Fig.~4a, the sine wave passes through zero at $R=8$~kpc.
Therefore, despite its significant amplitude $(f_R),$ the radial
component of the wave causes no shift of the mean velocities (its
main contribution to the velocity $U_{\odot LSR}).$ The amplitude
of the azimuthal component of the spiral wave $(f_\theta)$ is
almost zero (its main contribution to the velocity
${V_\odot}_{LSR}$). Then,
$(U_\odot,V_\odot,W_\odot)_{LSR}=(u_\odot,v_\odot,w_\odot)=(8.9,10.3,6.8)\pm(0.6,1.0,0.4)$~km
s$^{-1}$ according to solution 1 from Table 2.

The present-day situation with the determination of this velocity
by various methods but without invoking any data on young stars is
as follows. Sch\"{o}nrich et al. (2010) took into account the
stellar metallicity gradient in the Galactic disk and found the
following components:
 $(U_\odot,V_\odot,W_\odot)_{LSR}= (11.1,12.2,7.3)\pm(0.7,0.5,0.4)$ km s$^{-1}$.
Having analyzed the eccentricities of stars in the solar
neighborhood, Francis and Anderson (2009) found
 $(U_\odot,V_\odot,W_\odot)_{LSR}= (7.5,13.5.2,6.8)\pm(1.0,0.3,0.1)$ km s$^{-1}$.
Based on an updated version of the Geneva--Copenhagen survey
(Holmberg et al. 2007), Koval' et al. (2009) found
 $(U_\odot,V_\odot,W_\odot)_{LSR}= (5.1,7.9,7.7)\pm(0.4,0.5,0.2)$ km s$^{-1}$ from stars born at a
circumsolar distance by taking into account the radial migration
of stars and the metallicity gradient in the Galactic disk. Thus,
the solar peculiar velocity components are
 $(U_\odot,V_\odot,W_\odot)_{LSR}=(10,11,7)$ km s$^{-1}$.
The components we found from OB3 stars are in good agreement with
these values.

This is confirmed by the analysis of a sample of blue supergiants
that are closest in evolutionary status to the OB3 stars of our
sample:
 $(U_\odot,V_\odot,W_\odot)_{LSR}=(6,11,7)$ km s$^{-1}$ (Zabolotskikh et al. 2002),
 where the phase was $\chi_\odot=-97^\circ$ at a significant amplitude
 $f_R = -6.6\pm2.5$ km s$^{-1}$. A similar result was obtained by
Fern\'andez et al. (2001) both using a sample of OB stars,
 $(U_\odot,V_\odot,W_\odot)_{LSR}= (8.8, 12.4, 8.4)\pm(0.7, 1.0, 0.5)$ km s$^{-1}$, and from Cepheids,
 $(U_\odot,V_\odot,W_\odot)_{LSR}= (6.5, 10.4, 5.7)\pm(1.2, 1.9, 0.7)$ km s$^{-1}$, with a
phase $\chi_\odot\approx-30^\circ,$ although the spiral wave
amplitudes turned out to be insignificant.

(2) The ratio of the spiral wave amplitudes we found from our
sample of 102 OB3 stars,
 $f_R = -13\pm1$~km s$^{-1}$ and
 $f_\theta = 0\pm1$~km s$^{-1}$, is in good agreement with the analysis of blue supergiants by
Zabolotskikh et al. (2002),
 $f_R = -6.6\pm2.5$~km s$^{-1}$ and
 $f_\theta=0.4\pm2.3$~km s$^{-1}$, and a sample of Cepheids with close values of
these parameters.

At the same time, it can be seen from Fig. 4b that a wave with a
significant amplitude ($\approx$10 km s$^{-1}$) but with a
wavelength $\lambda\approx1$~kpc is present in the azimuthal
velocities. As we see from the figure, the local standard of rest
turns out to be shifted in the region $R\rightarrow R_0$ in the
direction opposite to the Galactic rotation. When a sample of
masers was analyzed (Bobylev and Bajkova 2010; Stepanishchev and
Bobylev 2011), a special allowance for the perturbations from the
spiral wave in the region $R\rightarrow R_0$ made it possible to
estimate the velocity components
$(U_\odot,V_\odot,W_\odot)_{LSR}.$

(3) The parameters of the Galactic rotation curve $\Omega_0$ and
$\Omega'_0$ calculated from 102 OB3 stars using the refined
``calcium'' distance scale (solutions 1,2,3, and 5 from Table 2)
are in good agreement with the results of the analysis of Cepheids
and blue supergiants (Zabolotskikh et al. 2002), OB associations
(Mel'nik and Dambis 2009), and masers (Bobylev and Bajkova 2010;
Stepanishchev and Bobylev 2011).

\section*{CONCLUSIONS}

We tested the distances derived from the equivalent widths of
interstellar CaII spectral lines by Megier et al. (2009). For this
purpose, we used a sample of 102 relatively distantOB3 stars with
known proper motions and radial velocities.

The internal reconciliation of the distance scales using the first
derivative of the angular velocity of Galactic rotation
$\Omega'_0$ and the external reconciliation with Humphreys's
distance scale for OB associations refined by Mel'nik and Dambis
(2009), we showed that the initial distances should be reduced by
$\approx$20\%.

In the refined distance scale, our OB3 stars are located at
heliocentric distances in the range from 0.6 to 2.6 kpc. We used
them to construct the Galactic rotation curve and to determine the
parameters of the spiral density wave.

For a fixed distance to the Galactic center, $R_0=8$ kpc, we found
the solar peculiar velocity components
 $(u_\odot,v_\odot,w_\odot)=(8.9,10.3,6.8)\pm(0.6,1.0,0.4)$~km s$^{-1}$; the angular velocity of
Galactic rotation
 $\Omega_0=-31.5\pm0.9$~km s$^{-1}$ kpc$^{-1}$ and its
derivatives
 $\Omega'_0= +4.49\pm0.12$ km s$^{-1}$ kpc$^{-2}$,
 $\Omega''_0= -1.05\pm0.38$ km s$^{-1}$ kpc$^{-3}$. The corresponding Oort constants are
 $A= 17.9\pm0.5$ km s$^{-1}$ kpc$^{-1}$ and
 $B=-13.6\pm1.0$ km s$^{-1}$ kpc$^{-1}$;
 the circular rotation velocity of the solar neighborhood is
 $|V_0|=252\pm14$~km s$^{-1}$;
 the amplitudes of the spiral density wave are
 $f_R=-12.5\pm1.1$~km s$^{-1}$ and
 $f_\theta=2.0\pm1.6$~km s$^{-1}$;
 the pitch angle of the two-armed spiral pattern is
 $i=-5.3^\circ\pm0.3^\circ$ and the phase
of the Sun in the spiral wave is
 $\chi_\odot=-91^\circ\pm4^\circ;$ and the
direction to the Galactic center is
 $l_0=-1.5^\circ\pm1.3^\circ.$ The
wavelength of the spiral density wave at the solar distance is
 $\lambda= 2.3\pm0.2$~kpc. It is particularly pronounced in the
Galactocentric radial velocities $V_R.$ The residual azimuthal
velocities $\Delta V_\theta$ have a more complex structure.

\bigskip{\bf ACKNOWLEDGMENTS}\bigskip

We are grateful to the referees for valuable remarks that
contributed to a significant improvement of the paper. The SIMBAD
search database provided a great help to our study. This work was
supported by the Russian Foundation for Basic Research (project
no. 08-02-0040) and in part by the ``Origin and Evolution of Stars
and Galaxies'' Program of the Presidium of the Russian Academy of
Sciences and the Program of State Support for Leading Scientific
Schools of the Russian Federation (project. NSh--3645.2010.2,
``Multiwavelength Astrophysical Studies'').

\bigskip{\bf REFERENCES}\bigskip
{\small

 1. V.S. Avedisova, Astron. Rep. 82, 488 (2005).

 2. V.V. Bobylev and A.T. Bajkova, MNRAS 408, 1788 (2010).

 3. V.V. Bobylev, A.T. Bajkova, and S.V. Lebedeva, Astron. Lett. 33, 720 (2007).

 4. V.V. Bobylev, A.T. Bajkova, and A.S. Stepanishchev, Astron. Lett. 34, 515 (2008).

 5. W.B. Burton, Astron. Astrophys. 10, 76 (1971).

 6. D. P. Clemens, Astrophys. J. 295, 422 (1985).

 7. Z. Cvetkovi\'c, I. Vince, and S. Ninkovi\'c , New Astron. 15, 302 (2010).

 8. A.K. Dambis, A.M. Mel'nik, and A.S. Rastorguev, Astron. Lett. 27, 58 (2001).

 9. W. Dehnen and J.J. Binney, MNRAS 298, 387 (1998).

 10. D. Fern\'andez, F. Figueras, and J. Torra, Astron. Astrophys. 372, 833 (2001).

 11. M. Fich, L. Blitz, and A.A. Stark, Astrophys. J. 342, 272 (1989).

 12. T. Foster and B. Cooper, astro-ph:1009.3220 (2010).

 13. C. Francis and E. Anderson, New Astron. 14, 615 (2009).

 14. A.M. Ghez, S. Salim, N.N. Weinberg, et al., Astrophys. J. 689, 1044 (2008).

 15. S. Gillessen, F. Eisenhauer, S. Trippe, et al., Astrophys. J. 692, 1075 (2009).

 16. J. Holmberg, B. Nordstr\"{o}m, and J. Andersen, Astron.Astrophys. 475, 519 (2007).

 17. R.M. Humphreys, Astrophys. J. Suppl. 38, 309 (1978).

 18. N.V. Kharchenko, R.-D. Scholz, A.E. Piskunov, et al., Astron. Nachr. 328, 889 (2007).

 19. V.V. Koval', V.A. Marsakov, and T.B. Borkova, Astron. Rep. 53, 1117 (2009).

 20. F. van Leeuwen, Astron. Astrophys. 474, 653 (2007).

 21. C.C. Lin, C. Yuan, and F.H. Shu, Astrophys. J. 155, 721 (1969).

 22. L. Mahy, G. Rauw, F. Martins, et al., Astrophys. J. 708, 1537 (2010).

 23. J. Maiz-Apell\'aniz, E.J. Alfaro, and A. Sota, astroph: 0804.2553 (2008).

 24. P.J. McMillan and J.J. Binney, MNRAS 402, 934 (2010).

 25. A. Megier, A. Strobel, A. Bondar, et al., Astrophys. J. 634, 451 (2005).

 26. A. Megier, A. Strobel, G.A. Galazutdinov, et al., Astron. Astrophys. 507, 833 (2009).

 27. A.M. Mel'nik and A.K. Dambis, MNRAS 400, 518 (2009).

 28. A.M. Mel'nik, A.K. Dambis, and A.S. Rastorguev, Astron. Lett. 27, 521 (2001).

 29. Yu.N. Mishurov and I.A. Zenina, Astron. Astrophys. 341, 81 (1999).

 30. I.I. Nikiforov, ASP Conf. Ser. 316, 199 (2004).

 31. K.F. Ogorodnikov, Dynamics of Stellar Systems (Pergamon, Oxford, 1965).

 32. D. Pourbaix, A.A. Tokovinin, A.H. Batten, et al., Astron. Astrophys. 424, 727 (2004).

 33. A.S. Rastorguev, E.V. Glushkova, A.K. Dambis, et al., Astron. Lett. 25, 595 (1999).

 34. M.J. Reid, Ann. Rev. Astron. Astrophys. 31, 345 (1993).

 35. M.J. Reid, K.M. Menten, X.W. Zheng, et al., Astrophys. J. 700, 137 (2009).

 36. K. Rohlfs, Lectures on Density Wave Theory (Springer, Berlin, 1977).

 37. R. Sch\"{o}nrich, J. Binney, and W. Dehnen, MNRAS 403, 1829 (2010).

 38. A.S. Stepanishchev and V.V. Bobylev, Astron. Lett. 37, 254 (2011).

 39. W.J. Tango, J. Davis, A.P. Jacob, et al., MNRAS 396, 842 (2009).

 40. N. Tetzlaff, R. Neuh\"{a}user, and M.M. Hohle, MNRAS 410, 190 (2011).

 41. The HIPPARCOS and Tycho Catalogues, ESA SP--1200 (1997).

 42. B.Y. Welsh, R. Lallement, J.-L.Vergely, et al., Astron. Astrophys. 510, A54 (2010).

 43. M.V. Zabolotskikh, A.S. Rastorguev, and A.K. Dambis, Astron. Lett. 28, 454 (2002).

Translated by N. Samus' }

\end{document}